\documentclass[prd,reprint,nofootinbib,superscriptaddress,preprintnumbers,showpacs,ctexart,article]{revtex4-1}
\usepackage{amsfonts}
\usepackage{amssymb}
\usepackage{amsmath}
\usepackage{graphicx,color}
\usepackage{dcolumn}
\usepackage{epsfig}
\usepackage{bm}
\usepackage{bbm}
\usepackage{ulem}
\usepackage{hyperref}
\usepackage{lineno}
\usepackage{float}
\usepackage{subfigure}
\usepackage{booktabs}
\usepackage{multirow}
\usepackage{siunitx}
\usepackage{array}
\allowdisplaybreaks[4]

\begin{document}
	
\title{Probing double hadron resonance by complex momentum representation method}

\author{Wen-Wan He}
\affiliation{School of Physics and Optoelectronics Engineering, Anhui University, Hefei 230601, People's Republic of China }
\author{Mao Song}	
\email{songmao@ahu.edu.cn}
\author{Jian-You Guo}
\affiliation{School of Physics and Optoelectronics Engineering, Anhui University, Hefei 230601, People's Republic of China }
\author{Xuan Luo}
\affiliation{School of Physics and Optoelectronics Engineering, Anhui University, Hefei 230601, People's Republic of China }
\author{Gang Li}
\affiliation{School of Physics and Optoelectronics Engineering, Anhui University, Hefei 230601, People's Republic of China }

\begin{abstract}
\vspace{0.5cm}
Resonances are ubiquitous phenomena in nature, and physicists have developed many methods to explore resonant states. Of particular note is the complex momentum representation(CMR) method, which has been developed and widely used in the study of resonant states in atomic, molecular and nuclear physics. Here, for the first time, we have developed this novel method to study hadron resonant states. The CMR method is applied to probe the bound and resonant states for the $\Lambda_cD(\bar{D})$ and $\Lambda_c\Lambda_c(\bar{\Lambda}_c)$ systems, in which the resonant states are exposed clearly in complex momentum plane and the resonance parameters can be determined precisely without imposing unphysical parameters. The results show that the CMR method has achieving higher accuracy than other widely used methods. This method is not only very effective for narrow resonances, but also can be reliably applied to broad resonances.

\end{abstract}
	\maketitle	
	\section{Introduction}
	\label{I}
Hadrons are composite subatomic particles made of two or more quarks held together by the strong interaction\cite{Gell-Mann:1964ewy,Zweig:1964jf,Fritzsch:1973pi}, which can be described by Quantum Chromodynamics (QCD) in the Standard Model. The quark model divides hadrons into two categories: one consisting of a quark-antiquark pair, called mesons; and the other consisting of three quarks, called baryons. Over the past years, these traditional hadrons predicted by quark model have been extensively studied experimentally and theoretically.
However, the QCD theory does not exclude other colorless hadron structures. These hardons meet the requirements of QCD color confinement, but have a different composition from the traditional hadrons, which are called exotic hadrons, such as $X(3872)$ \cite{Belle:2003nnu}, $Y(4260)$/$Y(4360)$ \cite{BaBar:2005xmz,BaBar:2006ait}, $Z_b(10610)/Z_b(10650)$ \cite{2012Observation}, $P_c(4380)/P_c(4450)$\cite{LHCb:2015yax}.

Theorists have proposed various models to explain the exotic states, such as the chiral effective field theory~\cite{Xu:2017tsr,Ohkoda:2012hv,Li:2012cs,Ren:2021dsi}, Bethe-Salpeter approach~\cite{Sakai:2017avl,He:2014nya,He:2015mja,Wallbott:2019dng}, constituent quark models~\cite{Zhu:2019iwm,Ortega:2021yis,Tan:2020ldi,Luo:2017eub}, QCD sum rules~\cite{Navarra:2007yw,Xin:2021wcr,Tang:2019nwv}, and relativized quark models~\cite{Lu:2020rog,Ebert:2007rn,Wang:2018pwi}, $etc$.
Since many of these exotic states $X/Y/Z/P_c$ are close to the thresholds of two hadrons, they are naturally considered as candidates of the hadronic molecules.
Notably, deuterium is the only stable hadronic molecular state discovered until now, which can be effectively explained within the framework of the one-boson-exchange(OBE) model\cite{Nils1994On,Nils1993From}. Based on this understanding, we assume that deuteron-like hadronic states present similar structures and interaction potentials. Over the past decade, hadronic molecular states have been employed to explain the structure and decay of many exotic hadron states, particularly following the discovery of charmonium-like $X/Y/Z$ and $P_c(4380)/P_c(4450)$ state\cite{2008Isa,Thomas:2008ja,2009XA,2009Y,2011Z,2015Identifying}. If these hadrons can form hadron bound states, they may also form resonant states with high angular momentum, which remains relatively under explored in hadronic physics.

Resonance phenomena are widely present in nature and prevalent across diverse fields such as atomic physics, molecular physics, nuclear physics, and chemical reactions, which is one of the most intriguing areas of contemporary research. Therefore, a series of methods for resonance has been proposed, including $R$-matrix method\cite{E1947Higher,1987Pole}, $K$-matrix method\cite{1991Level}, $J$-matrix method, scattering phase shift method\cite{Heller:1974zzc}, continuous spectrum theory, and coupling channel method. Considering the formidable challenge of solving scattering problems, several bound-state-like methods have been developed, which include the real stabilization method (RSM)\cite{1970Stabilization} and analytic continuation method of coupling constant (ACCC)\cite{1989Theory}.

However, there is a pressing need for a unified methodology capable of describing bound state, resonant state and continuum. The complex scaling method(CSM)\cite{1983The,1998Quantum} has satisfied this requirement and is extensively employed in exploring resonances within atomic, molecular, and nuclear physics. These advantages allow the CSM applying to different theoretical frameworks, including combinations with the few-body model\cite{2006Resonance}, shell model\cite{2008TOPICAL,2014Nuclear}, and Hartree-Fock theory\cite{2008Gamow,1997Particle}. In Ref.\cite{Yu:2021lmb}, we applied CSM to investigate the resonant states in hadronic physics. When applied to $Y(4630)$, which is explained successfully as a resonant state of the $\Lambda_c \bar{\Lambda}_c$ system\cite{Mei:2022msh}. Recently, CSM have been widely used in hadronic physics~\cite{Wang:2022yes,Cheng:2022qcm,Wang:2023ivd}. Nevertheless, the CSM is not without its limitations. In complex scaling calculations, multiple diagonalizations of the Hamiltonian are required to accurately identify resonance parameters. Furthermore, unphysical parameters, such as the complex rotation angle, also have an impact on the calculation results.

To address these challenges, a complex momentum representation(CMR) method has been established for resonant states\cite{1968On,Deltuva2015Momentum,2006Study}, which represents the Schr\"{o}dinger equation in momentum space. This method retains the advantages of the CSM, and is convenient to deal with both bound and resonant states consistently\cite{Sukumar:1978rf,Kwan:1978zh,Li:2016gbp}. Compared to the CSM, CMR method does not need to adjust unphysical parameters, which improves both convenience and calculation precision. Furthermore, CMR is particularly effective at calculating wave function and density of the bound state. In view of these advantages, we will adopt the CMR method in hadronic physics to identify the resonant states.

In heavy molecular state systems\cite{2009X}, it is essential to consider the contributions from $\pi$, $\eta$, $\rho$, $\sigma$, $\omega$ mesons, notably, one-$\pi$ meson exchange exerts a dominant influence. For $\Lambda_cD(\bar{D})$ and $\Lambda_c\Lambda_c(\bar{\Lambda}_c)$ systems, the contribution of $\pi$, $\eta$, $\rho$ meson exchanges are severely prohibited or inhibited, the contributions of one-$\sigma$-exchange(OSE) and one-$\omega$-exchange(OOE) are dominant\cite{2017Prediction}. In the CSM, we have studied the possible bound states and resonant states of these systems\cite{Yu:2021lmb}. In this paper, we will employ the CMR method to calculate the bound and resonant states for the $\Lambda_cD(\bar{D})$ and $\Lambda_c\Lambda_c(\bar{\Lambda}_c)$ systems\cite{2011PossibleL,2011Possibled}. The calculation results are compared with those by the CSM.

This paper is organized as follows. In Sec. \ref{sec2} the theoretical framework and the CMR method are illustrated. Sec. \ref{sec3} gives the numerical results and discussion. A short conclusion is given in Section \ref{sec4}.

\section{Framework}\label{sec2}
\label{II}
In order to better verify the feasibility and validity of the CMR method, we chose a relatively simple hadron system $\Lambda_cD(\bar{D})$ and $\Lambda_c\Lambda_c(\bar{\Lambda}_c)$ for investigation.
Based on the spin and isospin conservation, there is no coupling $\pi \Lambda_c \Lambda_c$ and $\pi D D$, and the contributions of $\pi$, $\eta$, $\rho$ meson exchanges are severely prohibited or inhibited. The effective Lagrangians for one-$\sigma$-exchange and one-$\omega$-exchange can be expressed under the heavy-quark symmetry\cite{2017Prediction},
\begin{eqnarray}
\mathcal{L}_{DD\sigma/\omega} &=& -2g_{\sigma}D^{\dag}D\sigma+2g_{\omega} D^{\dag}D \bm{\upsilon} \cdot \omega,\label{DD}\\
\mathcal{L}_{\Lambda_c\Lambda_c\sigma/\omega} &=& -2g_{\sigma}'\bar{\Lambda}_c\Lambda_c\sigma-2g_{\omega}'\bar{\Lambda}_c\Lambda_c \bm{\upsilon} \cdot \omega.\label{LL}
\end{eqnarray}
Here, $\bm{\upsilon}$ is the four velocity of the heavy meson, which has the form of $\bm{\upsilon}=(1,\vec{0})$.

According to the effective Lagrangians in Eqs. (\ref{DD}) and (\ref{LL}), the relevant OBE scattering amplitudes can be collected in Table \ref{no1}.
\begin{table}[htbp]
  \centering
  \caption{The scattering amplitudes for all investigated systems. Here, $\mathcal{H}(\vec{q},m)$ is defined as $\mathcal{H}(\vec{q},m)=1/(\vec{q}^2+m^2)$.}\label{no1}
  \begin{tabular}{cc}
    \toprule[2pt]
    $h_1h_2\to h_3h_4$  & $\mathcal{M}(h_1h_2\to h_3h_4)$ \\\hline
    \midrule
  $\Lambda_c\bar{D}\to \Lambda_c\bar{D}$
                   &$8M_{D}M_{\Lambda_c}\chi_3^{\dag}\chi_1\left[g_{\sigma}g_{\sigma}'\mathcal{H}(\vec{q},m_{\sigma})-g_{\omega}g_{\omega}'\mathcal{H}(\vec{q},m_{\omega})\right]$\\
  $\Lambda_c{D}\to \Lambda_c{D}$
                   &$8M_{D}M_{\Lambda_c}\chi_3^{\dag}\chi_1\left[g_{\sigma}g_{\sigma}'\mathcal{H}(\vec{q},m_{\sigma})+g_{\omega}g_{\omega}'\mathcal{H}(\vec{q},m_{\omega})\right]$\\
  $\Lambda_c\Lambda_c\to\Lambda_c\Lambda_c$
       &$16M_{\Lambda_c}^2\chi_3^{\dag}\chi_4^{\dag}\chi_1\chi_2\left[g_{\sigma}g_{\sigma}'\mathcal{H}(\vec{q},m_{\sigma})-g_{\omega}g_{\omega}'\mathcal{H}(\vec{q},m_{\omega})\right]$\\
  $\Lambda_c\bar{\Lambda}_c\to\Lambda_c\bar{\Lambda}_c$
       &$16M_{\Lambda_c}^2\chi_3^{\dag}\chi_4^{\dag}\chi_1\chi_2\left[g_{\sigma}g_{\sigma}'\mathcal{H}(\vec{q},m_{\sigma})+g_{\omega}g_{\omega}'\mathcal{H}(\vec{q},m_{\omega})\right]$\\
    \bottomrule[2pt]
  \end{tabular}
\end{table}

With the Born-Oppenheimer approximation, we can obtain that the relationship between the scattering amplitude in the momentum space and the effective potential in the momentum space is
\begin{eqnarray}
\mathcal{V}_{fi}(\vec{q}) &=& -\frac{\mathcal{M}_{fi}(h_1h_2\to h_3h_4)}{\sqrt{\prod_i2m_i\prod_f2m_f}}.
\end{eqnarray}
Here, the scattering amplitude for the process $h_1h_2\to h_3h_4$ is $\mathcal{M}(h_1h_2\to h_3h_4)$. $m_i$ and $m_f$ are the masses of the initial and final particles, respectively.

In order to regularize the off shell effect of the exchanged mesons and the structure effect of the hadrons, a monopole form factor $\mathcal{F}(q^2)$ is introduced at every vertex,
\begin{eqnarray}
\mathcal{F}(q^2) &=& \frac{\Lambda^2-m^2}{\Lambda^2-q^2}.
\end{eqnarray}
Here, $\Lambda$ is the cut-off parameter, $m$ correspond to the mass of exchanged meson, the four momentum of exchanged meson $q$ is $q^\mu = (0, \vec{q})$. $\Lambda$ is related to the root-mean-square radius of the source hadron which is usually chosen to be around 1.0 GeV.

Then, we obtained the interaction potential in momentum space as shown in Table \ref{no2}. Here, we define the function $Y(\vec{q},\mathrm{m})$ as
\begin{equation}
Y(\vec{q},\mathrm{m})=\mathcal{H}(\vec{q},m)\mathcal{F}^2(\vec{q}^2).
\end{equation}

\begin{table}[htbp]
  \centering 
  \caption{The effective potentials in momentum space for all investigated systems.}\label{no2} 
  \begin{tabular}{ccc} 
    \toprule[2pt] 
    Systems & Quarks & $V(\vec{q})$ \\\hline
    \midrule 
    $\Lambda_{c}\bar{D}$ & $(cqq)(\bar{c}{q})$ & $-2g_{\sigma}g_{\sigma}'Y(\vec{q},m_{\sigma})+2g_{\omega}g_{\omega}'Y(\vec{q},m_{\omega})$ \\
    $\Lambda_c{D}$ & $(cqq)(c\bar{q})$ & $-2g_{\sigma}g_{\sigma}'Y(\vec{q},m_{\sigma})-2g_{\omega}g_{\omega}'Y(\vec{q},m_{\omega})$ \\
    $\Lambda_c\Lambda_c$ & $(cqq)(cqq)$ & $-4g_\sigma'^{2}Y(\vec{q},m_{\sigma})+4g_\omega'^{2}Y(\vec{q},m_{\omega})$ \\
    $\Lambda_c\bar{\Lambda}_c$ & $(cqq)(\bar{c}\bar{q}\bar{q})$ & $-4g_\sigma'^{2}Y(\vec{q},m_{\sigma})-4g_\omega'^{2}Y(\vec{q},m_{\omega})$ \\
    \bottomrule[2pt] 
  \end{tabular}
\end{table}

For a nonrelativistic system, the Schr\"{o}dinger equation can be represented as
\begin{equation}
H|\psi\rangle=E|\psi\rangle,\label{Hamiltonian}
\end{equation}%
where the $|\psi\rangle$ is wavefunction and the $H$ is the Hamiltonian operator
\begin{equation}
H=T+V,
\end{equation}%
the kinetic energy operator, denoted as $T$, is given by $T=\frac{\vec{q}^{2}}{2\mu}$ , where $\mu = m_1 m_2/(m_1 +m_2)$ denotes
the reduced mass of two-body system and $\vec{q}$ is the relative three momentum, which is equal to the wavevector $\vec{k}$ in natural units.
The solutions of Eq.~(\ref{Hamiltonian}) encompass the spectra of bound states, resonant states, and the continuum.
The bound state can be obtained using standard methods. For the resonant states, the momentum representation is employed with the Schr\"{o}dinger equation:
\begin{equation}
\int d\vec{k}^{\prime}\langle \vec{k}|H|\vec{k}^{\prime}\rangle \psi(\vec{k}^{\prime})=E\psi (\vec{k}), \label{three-Schrodinger}
\end{equation}%

By substituting equation Eq.~(\ref{Hamiltonian}) into equation Eq.~(\ref{three-Schrodinger}), the potential in momentum space can be expanded by partial-wave method,

\begin{equation}
\frac{\hbar^2k^2}{2\mu}\psi(\vec{k})+\int \sum_{l}\frac{1}{(2\pi)^{3}} V_l\left(k,k'\right)P_l(\cos\theta)\psi(\vec{k}')d\vec{k}'=E\psi(\vec{k}).\label{three-Schrodinger2}
\end{equation}%
Here, the exchanged three momentum $\vec{q} = \vec{k}-\vec{k'}$, and $\psi(\vec{k})$ represents the momentum wavefunction, with wavevector $\vec{k}$ corresponding to momentum.  The Schr\"{o}dinger equation has become an integral equation in momentum space. $V_l(k,k')$ is the partial-wave expansion potential,
with
\begin{equation}
V_l(k,k')=\frac{2l+1}{2}\int_{-1}^1 V(\vec{q})P_l(\cos\theta)d\cos\theta.
\end{equation}

Instead of solving the three-dimensional integral equation given in
Eq.~(\ref{three-Schrodinger}), an infinite set of one-dimensional equations can be obtained by invoking a partial wave expansion.
The wavefunction $\psi(\vec{k})$ can be expanded in a complete set of spherical harmonics as
\begin{eqnarray}
\psi\left(\vec{k}\right)=f^l\left(k\right)Y_{lm}\left(\Omega_k\right).
\end{eqnarray}
The legendre polynomials can be expressed as the addition of spherical harmonics,
\begin{eqnarray}
P_l(\cos\theta) = \frac{4\pi}{2l+1} \sum_{m=-l}^l Y_{lm}(\Omega_k)Y^*_{lm}(\Omega_{k'}).
\end{eqnarray}

After integrating over the solid angles, the Schr\"{o}dinger equation in momentum space becomes
\begin{equation}
\frac{\hbar^2k^2}{2\mu}f^l(k)+\int_0^\infty \frac{4\pi}{2l+1} \frac{1}{(2\pi)^{3}} V_l\left(k,k'\right)f^l(k')k'^2dk'=Ef^l(k),\label{Schrodinger}
\end{equation}%

This possesses numerous advantageous characteristics. Firstly, it explicitly provides the most realistic hadron-hadron interactions derived from effective field theory in momentum space. Secondly, the boundary conditions imposed on the differential equation in coordinate space are seamlessly incorporated into the integral equation. Lastly, but importantly, integral equations are straightforward to implement numerically, and convergence can be achieved simply by increasing the number of integration points.

In order to solve the Schr\"{o}dinger equation numerically, the Gaussian-Legendre quadrature method is used to handle the momentum space integral.
Since the Hamiltonian matrix is asymmetrical, we make it symmetrical by the transformation
\begin{equation}
  \mathbf{f}(k_a)=\sqrt{w_a}k_a f(k_a)
\end{equation}
The Schr\"{o}dinger equation becomes
\begin{equation}
\frac{\hbar^2k_a^2}{2\mu}\mathbf{f}^l(k_a)+\sum_b \frac{4\pi k_ak_b\sqrt{w_aw_b}}{(2l+1)(2\pi)^{3}} V_l(k_a,k_b)\mathbf{f}^l(k_b)=E \mathbf{f}^l(k_a).
\end{equation}

Then, we can get the real space radial wavefunction in the coordinate space by
\begin{equation}
 f^l\left(r\right)=\quad i^l \sqrt{\frac{2}{\pi}}\sum_{a=1}^N \sqrt{w_a}k_aj_l (k_ar)\mathbf{f}^l(k_a),
\end{equation}
the density in the coordinate space is,
\begin{equation}
 \rho^l(r)=f^{l*}(r)f^l(r).
\end{equation}

\section{Results}\label{sec3}
\label{III}
In this section, we discuss and analysis the effects of the OSE and OOE interactions for the systems of $\Lambda_cD (\bar{D})$ and  $\Lambda_c \Lambda_c (\bar \Lambda_c)$ by solving the Schr\"{o}dinger equation in the CMR. The related parameters required for this work are sourced from Ref~\cite{2016Review}, which are summarized in Table \ref{num1}.
\renewcommand\tabcolsep{0.16cm}
\renewcommand{\arraystretch}{1.8}
\begin{table}[!htbp]
  \caption{The related parameters are used in this work.}\label{num1}
  \begin{tabular}{ccc|ccc}\toprule[2pt]
  Hadron     &$I(J^P)$     &Mass (MeV)    &Hadron     &$I(J^P)$     &Mass (MeV) \\\hline
  $D$        &$\frac{1}{2}(0^-)$    &1867.24      &$\Lambda_c$     &$0(\frac{1}{2}^+)$     &2286.46\\
  $\sigma$   &$0(0^+)$              &600          &$\omega$        &$0(1^-)$               &782.65\\
  \bottomrule[2pt]
  \end{tabular}
\end{table}

The coupling strengths can be approximated utilizing the quark model, in which the $\sigma$ and $\omega$ mesons couple dominantly interact with light quarks in heavy hadrons.
According to this approximation, the coupling strengths of ${DD\sigma/\omega}$ and ${\Lambda_c\Lambda_c\sigma/\omega}$ are equal to the coupling strengths of ${qq\sigma/\omega}$.
In a $\sigma$ model\cite{1999Two}, the value of the coupling constant $g_{\sigma}$ is set to 3.65. As for the $\omega$ coupling constant $g_{\omega}$, the Nijmegen model provides a value of 3.45, whereas the Bonn model\cite{1999Soft} assigns it a value of 5.28. In Ref.\cite{2001Nucleon}, $g_{\omega}$ was roughly assumed to be 3.00. In the following calculations, all the possible values of the coupling constants will be taken into account.

\begin{figure}
  \centering
  \includegraphics[width=3.33in, keepaspectratio]{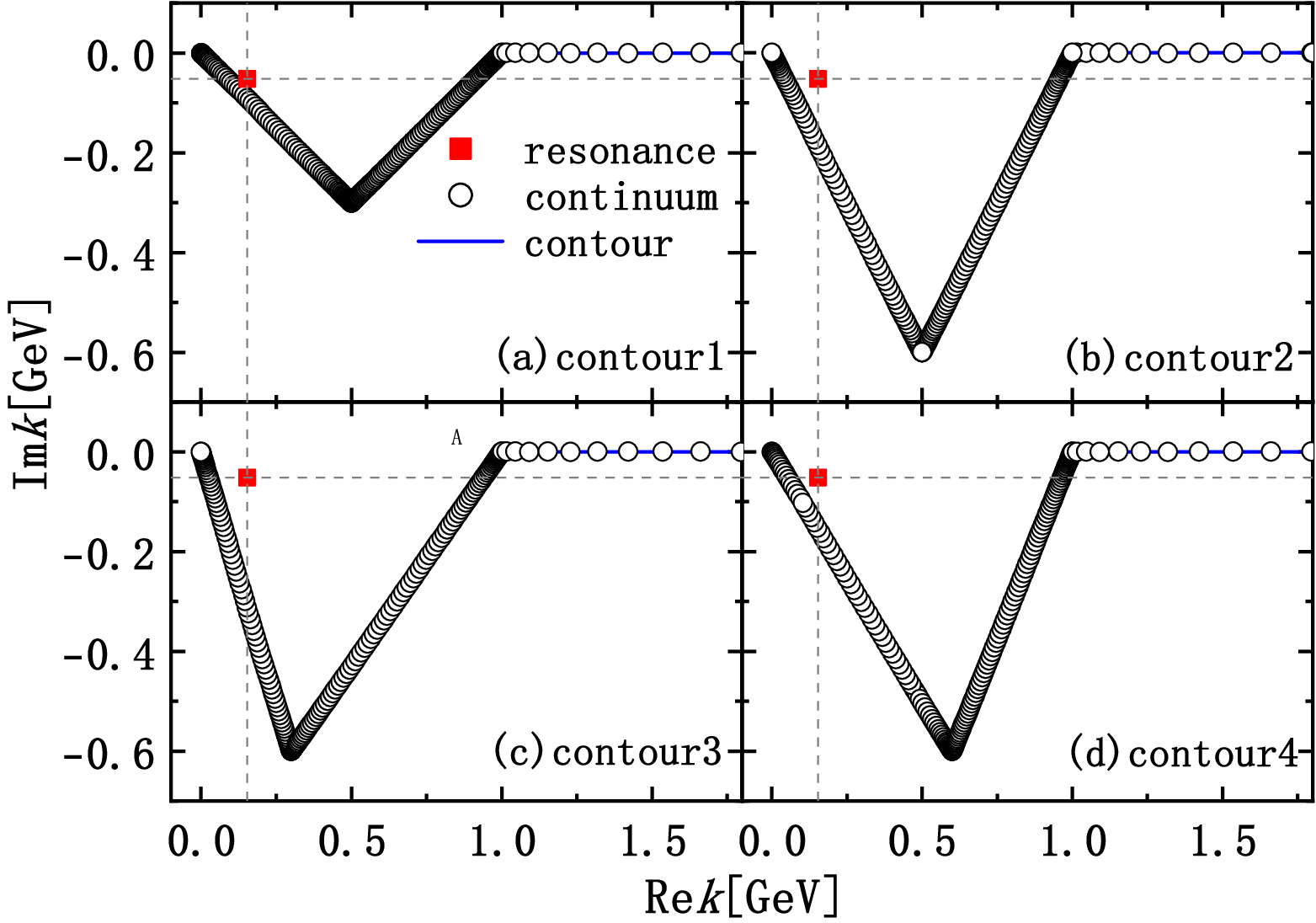}
  \caption{(Color online) The complex momentum solutions of the Schr\"{o}dinger equation for the $P$-wave resonant state with four different contours. The red filled square, black open circles and blue solid line in every subfigure represent the resonance, the continuum and the contour of integration in the complex momentum place, respectively.}
  \label{fig1}
\end{figure}

In order to pick out a suitable integration contour, we first need to check whether the current calculations depend on the choice of integration contour. In Fig.\ref{fig1}, we show the calculation of four different contours for the $P$-wave resonant state. It can be seen that the resonant state is separated from the continuum. In each subgraph, the exposed resonant state in the complex $k$ plane can be clearly seen. Compared to Fig.\ref{fig1}(a), the contour in Fig.\ref{fig1}(b) is deeper, and the continuum moves along the contour while the resonant state remains in its original position. Similarly, when the contour moves from right to left or from left to right, as shown in Fig.\ref{fig1}(c) and Fig.\ref{fig1}(d), the corresponding continuous spectra move with the contour, while the position of the resonant state does not change. This proves that the position of the resonant state is independent of the selection of integral contour. The resonant state is approximately located at the point with $k = 0.1538588 - i0.5204488$ $\rm{GeV}$ on the complex momentum plane in each subgraph. As shown in Fig.\ref{fig1}, although the contours are different, the position of the resonant state remains stationary when exposed on the plane of complex momentum.

\begin{figure}
  \centering
  \includegraphics[width=3.33in, keepaspectratio]{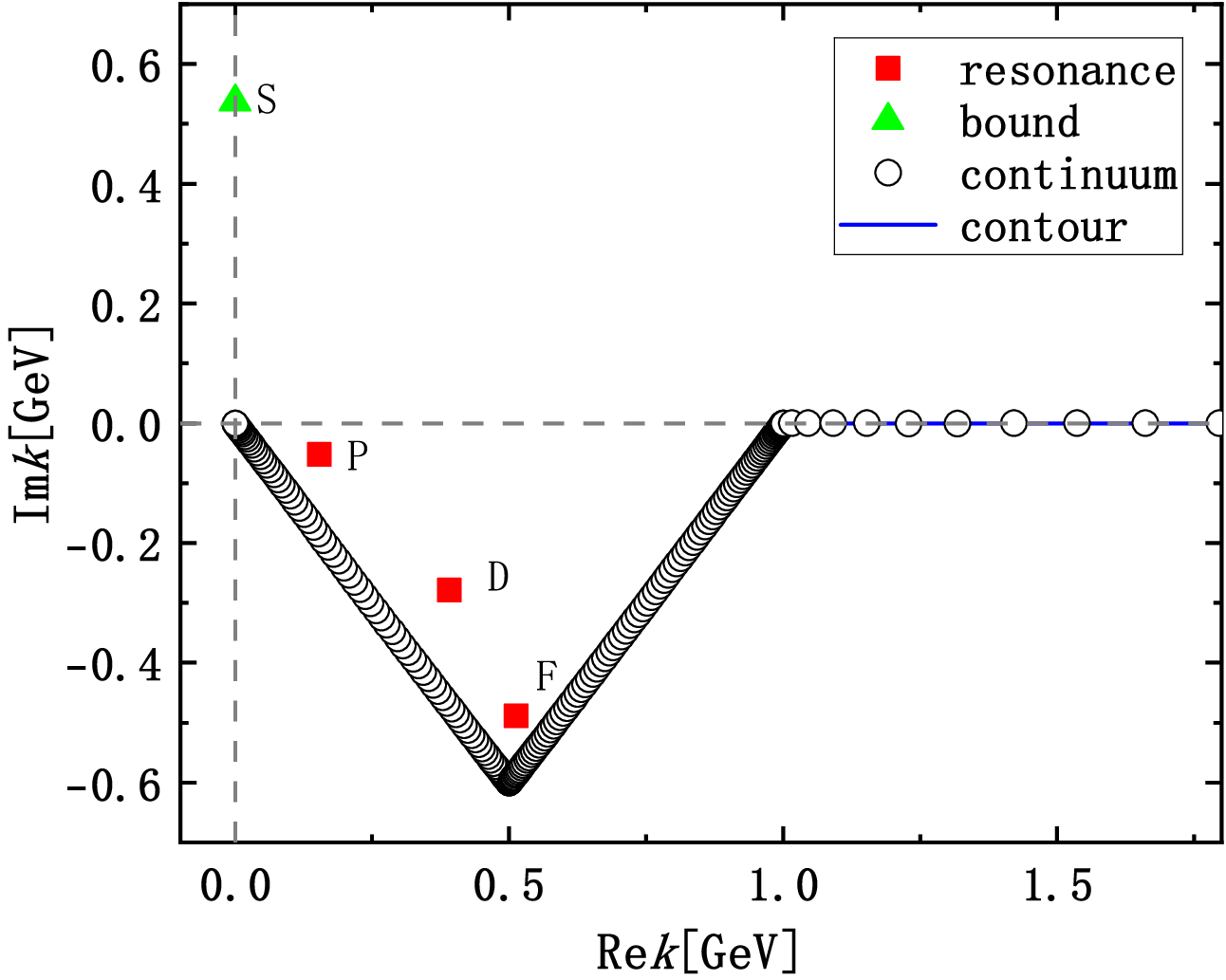}
  \caption{(Color online) The complex momentum solutions of the Schr\"{o}dinger equation for the $S$, $P$, $D$ and $F$-wave states. The red filled square, green filled triangular and black open circles represent the bound states, the resonant states, and the continuum, respectively. The blue solid line denotes the contour of integration in the complex momentum plane.}
  \label{fig2}
\end{figure}

The complex momentum solutions of the Schr\"{o}dinger equation for the $S$, $P$, $D$ and $F$-wave states are presented in Fig.\ref{fig2}. We can see that the resonant states are in the fourth quadrant, while all the bound states are on the imaginary axis of the complex momentum plane. Since the resonant state is independent of the contour, we can choose a contour that is large enough to expose all the associated resonances. The four points of the triangle profile we chose are $k = 0$ $\rm{GeV}$, $0.5 - i0.6$ $\rm{GeV}$, $1.0$ $\rm{GeV}$ and $k_{max} = 4.0$ $\rm{GeV}$.

\begin{table}[!htbp]
  \caption{The bound and resonant states for the $\Lambda_cD(\bar{D})$ and $\Lambda_c\Lambda_c(\bar{\Lambda}_c)$ systems. The energies of bound states $E$ are real. The resonance energies $E$ are complex as $E_r - i\Gamma/2$ where $E_r$ is the resonance energy and $\Gamma/2$ is the half-width. The cut-off parameter $\Lambda$ is set as 1.1~GeV. The value of coupling constant $g_{\omega}$ are set as $g_{\omega}=3.00$, 3.45 in the Nijmegen model, and 5.28 in the Bonn model, respectively. The notation $\ldots$ stands for no bound or resonant state solutions.}\label{num2}
  \resizebox{\linewidth}{!}{
  \begin{tabular}{c|ccccccc}\toprule[2pt]
     \multirow{2}*{\,$g_{\omega}$\,} & \multicolumn{2}{c}{$\Lambda_c\bar D$} & \multicolumn{1}{c}{$\Lambda_cD$} & \multicolumn{1}{c}{$\Lambda_c\Lambda_c$} & \multicolumn{1}{c}{$\Lambda_c\bar{\Lambda}_c$} \\
     \cmidrule{2-3} \cmidrule{5-5} \cmidrule{6-6} \cmidrule{7-7}

                        &L   &$E [\rm{MeV}]$                 &$E [\rm{MeV}]$           &$E [\rm{MeV}]$                 &$E [\rm{MeV}]$ \\\hline
  3.0                   &S   &\ldots                         &-11.094                  &-31.695                          &-110.388 \\
                        &P   &\ldots                         &\ldots                  &7.739 - i27.785                   &10.417 - i9.660\\
                        &D   &\ldots                         &\ldots                  &\ldots                            &28.666 - i98.113\\
                        &F   &\ldots                         &\ldots                  &\ldots                            &2.056 - i219.025\\\hline

  3.45                  &S   &\ldots                         &-14.406                  &-22.746                           &-126.019 \\
                        &P   &\ldots                         &0.272 - i46.513          &5.577 - i30.215                   &9.169 - i7.004\\
                        &D   &\ldots                         &\ldots                  &\ldots                            &32.925 - i95.097\\
                        &F   &\ldots                         &\ldots                  &\ldots                            &10.806 - i218.898\\\hline

  5.28                  &S   &\ldots                         & -38.661                 &\ldots                          &-222.567        \\
                        &P   &\ldots                         &9.643 - i37.763          &\ldots                          &\ldots         \\
                        &D   &\ldots                         &\ldots                  &\ldots                          &49.224 - i74.845\\
                        &F   &\ldots                         &\ldots                  &\ldots                          &52.463 - i208.839\\
  \bottomrule[2pt]
  \end{tabular}
  }
\end{table}

In Table \ref{num2}, we present the numerical results of bound and resonant states for the $\Lambda_c D (\bar D)$ and $\Lambda_c \Lambda_c (\bar \Lambda_c)$ systems with $\Lambda$ = 1.1~GeV. For these systems, the contributions of the mesons $\pi, \eta, \rho$ exchange are suppressed, and the intermediate- and short-range forces from $\sigma/\omega$ exchange contributions are dominant in Ref.\cite{2017Prediction}. The results show that $\Lambda_c \bar D$ system can not form either bound or resonant states. Both the $\Lambda_c D$ and $\Lambda_c \Lambda_c (\bar \Lambda_c)$ systems have resonant states over a wide range of parameters. Therefore, it is possible to compare the results of the $\Lambda_c \bar \Lambda_c$ system in the CSM and CMR method.
\begin{table}[!htbp]
  \centering
  \caption{The calculated results for the bound and resonant states in $\Lambda_c\bar{\Lambda}_c$ systems by using the CMR method in comparison with those obtained in coordinate representation by the CSM. Here, the energies of bound states $E$ are real, the resonance energies $E$ are complex as $E_r - i\Gamma/2$ where $E_r$ is the resonance energy and $\Gamma/2$ is the half-width, and the units is MeV. The cut-off parameter $\Lambda$ is set to 1.1~GeV, and the value of the coupling constant $g_{\omega}$ is set as $g_{\omega}=3.45$ in Nijmegen model.}\label{num3}
  \begin{tabular}{ccc}
    \toprule[2pt]
       &CMR  &CSM\\\hline
    L  & $E [\rm{MeV}]$ & $E [\rm{MeV}]$\\\hline
    \midrule
    S & -126.019 & -126.02 \\
    P & 9.169 - i7.004& 9.17 - i7.01 \\
    D & 32.925 - i95.097 & 32.92 - i95.10 \\
    F & 10.806 - i218.898 & 10.81 - i218.90 \\
    \bottomrule[2pt]
  \end{tabular}
\end{table}

Table \ref{num3} lists the numerical results of the bound and resonant states in Fig.\ref{fig2}. Compared them with the results obtained by the CSM in the coordinate space, we can see that these two results are in good agreement. This indicates that the results of the two methods in describing the bound state and resonant state are consistent. Nevertheless, the CMR results are more accurate than those obtained by the CSM method. Furthermore, the CMR method is not affected by any unphysical parameters.
\begin{figure}
  \centering
  \includegraphics[width=3.33in, keepaspectratio]{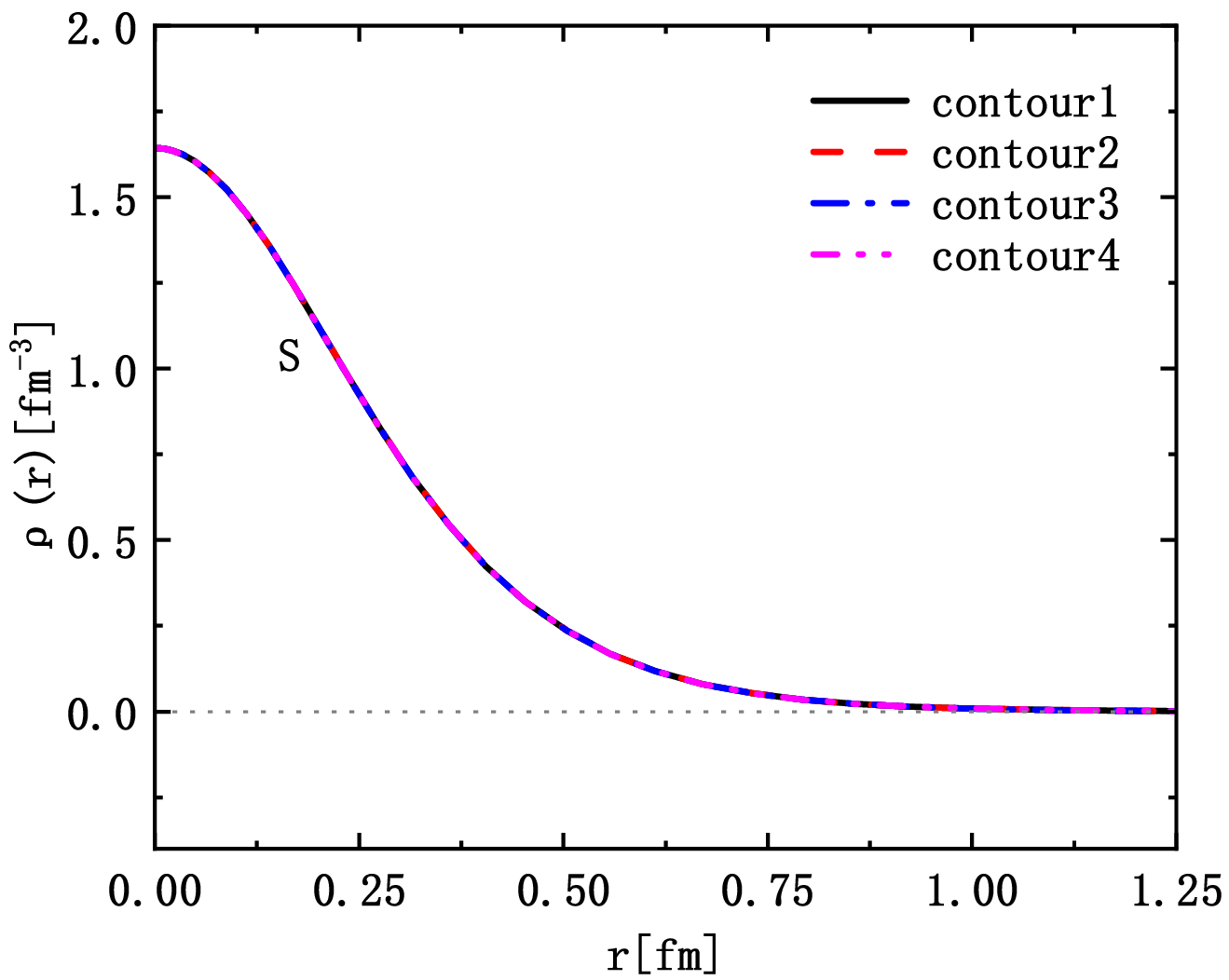}
  \caption{(Color online) The density distributions $\rho(r)$ of $S$ wave bound state in the coordinate space with four different contours for the momentum integration in the CMR calculations. The black solid, red dashed, blue dash-dotted, and magenta dash-dot-dotted lines represent the contour1, contour2, contour3, and contour4, respectively. The four contours are the same as those in Fig. \ref{fig1}.}
  \label{fig3}
\end{figure}
\begin{figure}
  \centering
  \includegraphics[width=3.33in, keepaspectratio]{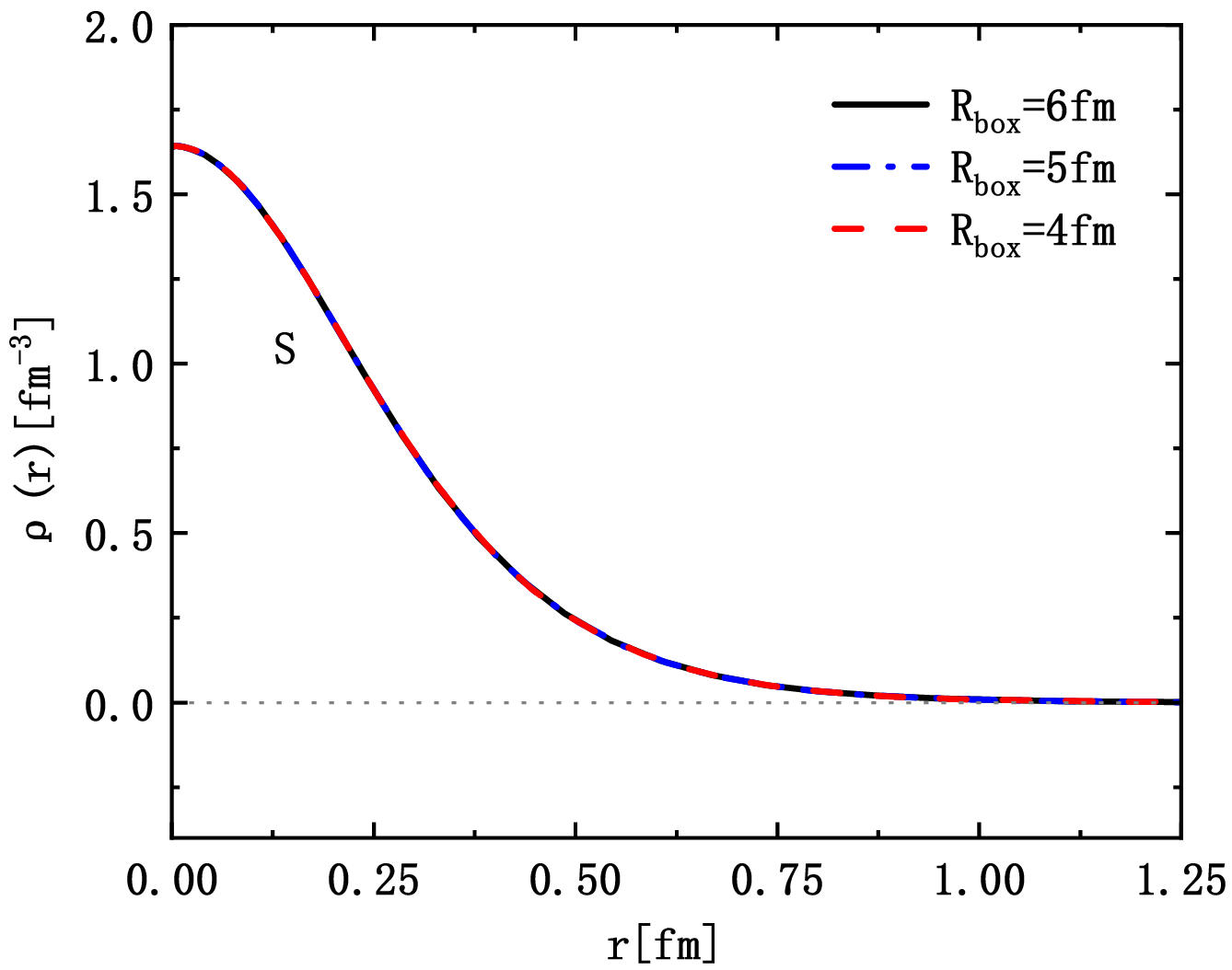}
  \caption{(Color online) The density distributions $\rho(r)$ for the S-wave bound state at energy $E = 126.01$ MeV are plotted in the coordinate space. The calculations are done with different space sizes, where the black solid, blue solid, and red solid represent $R_{box}$ = 4 fm, 5 fm, and 6 fm, respectively.}
  \label{fig4}
\end{figure}

The wave function and density of the bound state can be calculated by the CMR method. In the coordinate space, the densities of the bound state with four different contours are drawn in Fig.\ref{fig3}. We can see that the density distribution $\rho(r)$ of the bound state is also independent on the choice of the integration contours.

In Fig.\ref{fig4}, the density distribution $\rho(r)$ at the bound energy $E = 126.019$ MeV for the $S$-wave state is shown. The space dependence is also checked by doing calculations in different box sizes $R_{box}$ = 4 fm, 5 fm, and 6 fm. Exactly the same density distribution in the whole coordinate space is obtained with different space sizes.

\section{Conclusion}\label{sec4}
\label{IV}
In recent years, hadronic resonant state is one of the interesting topics of hadronic physics. The CSM is widely used in the calculation of hadronic resonant states. In order to overcome the drawbacks of the CSM, such as the results dependent on unphysical parameters and lower calculation accuracy, we extend the CMR method to describe bound and resonant states in hadronic physics here.  With $\Lambda_cD (\bar{D})$ and  $\Lambda_c \Lambda_c (\bar \Lambda_c)$ systems as illustrative examples, we have provided a detailed description to the theoretical formalism, and discussed the numerical results. We have checked the dependence of the calculated results on the contour and found that as long as exposed in the complex momentum plane, all the bound and resonant states are independent of the contours. We have also compared the results with that obtained by the CSM in coordinate space, and obtained consistent numerical results within the error range. In general, the CMR method can obtain more accurate calculation results and has a wider range of applications compared to the CSM.

\vfill
\section*{ACKNOWLEDGMENTS}
This work was supported in part by the National Natural Science Foundation of China (No.12475115, No.12205002, No.11935001), the Natural Science Foundation of Anhui Province (No.2208085MA10), and the Key Research Foundation of Education Ministry of Anhui Province of China (No.KJ2021A0061).


\end{document}